\documentclass[preprint]{aastex}
\begin{document}   
\title{Deprojecting Sunyaev-Zel'dovich statistics}
\normalsize
\author{Pengjie Zhang}
\affil{60 St.George Street, Astronomy Department, University of Toronto,
Canada, M5S 3H8}
\email{zhangpj@cita.utoronto.ca}
\and
\author{Ue-Li Pen}
\affil{60 St.George Street, Canadian Institute for Theoretical
Astrophysics, University of Toronto, Canada, M5S 3H8}
\email{pen@cita.utoronto.ca}
\begin{abstract}

We apply the hierarchical clustering model and non-linear perturbation
theory to the cosmological density and temperature fields.  This
allows us to calculate the intergalactic gas pressure power spectrum,
SZ anisotropy power spectrum, skewness and related statistics. Then we
show the effect of the non-gravitational heating. Our model confirms
recent simulations yielding mass weighted gas temperature $\bar{T}_g
\sim 0.35$ keV and reproduces the power spectra found in these simulations.

While the SZ effect contains only angular information, we show that it
is possible to extract the full time resolved gas pressure power
spectrum when combined with galaxy photometric redshift surveys by
using a variation on the cross-correlation. This method further
allows the disentanglement of the gravitational and the non-gravitational
heating.

\end{abstract} 
\section{Introduction}

One of the open questions in cosmology today is the state of the
intergalactic medium (IGM).  In the present popular cosmological
models, a putative cosmological constant accounts for perhaps 2/3 of
the energy density of the universe, and a similarly mysterious cold
dark matter component accounts for 2/3 of the remainder.  Of order
10\% of the energy density is supposed to be accounted for by baryons
\citep{Lange00,Tytler00} as obtained from primordial nucleosynthesis
calculation.  Surprisingly, the vast majority of the baryons in the
local universe is as of yet undetected.  The directly observed known
components consisting of stars, as well as hot and cold gas in
galaxies, groups and clusters, only account for a few percent of this
baryon budget \citep{Fukugita97}.  Cluster of galaxies are known to
possess a large gas fraction, of order $20\% h_{50}^{-3/2}$
\citep{Pen97}.  This gas fraction, if representative of the universe
at large, is sufficient to account for the majority of baryons in the
universe \citep{White93}.  Since gravity acts on all particles equally,
one expects this ratio within the virial radius of clusters to be
representative of the universal fraction.  In the standard bottom-up
picture of structure formation, clusters are the most massive objects
in the universe, and therefore the latest to form.  This suggests that
the majority of baryons is in a diffuse intergalactic form, which
forms the intracluster medium when clusters collapse.  If the baryons
were in other forms, for example compact objects which formed early in
the universe, it would be extremely difficult to understand how those
baryons would be released back into diffuse form when clusters form.

\citet{Pen99} showed that the IGM must have been heated by sources
other than the gravitational energy of collapse and its resulting
shock waves.  Its direct detection would be of significant interest to
quantitatively understand the distribution and state of the IGM, and
thereby infer its thermal history.  Unfortunately, hot diffuse gas is
difficult to detect in emission.  Luckily, the Cosmic Microwave
Background (CMB) scatters off all free electrons through inverse
compton scattering, allowing us to ``see'' the IGM through scattering,
in analogy to absorption.  This is known as the Sunyaev-Zel'dovich
effect \citep{Zeldovich69}, which is becoming routinely observable.
It becomes the dominant source of CMB fluctuations on arc-minute
scales.  Several experiments to conduct deep blank field or all sky
search are under development.  The upcoming CMB experiments such as
AMIBA (\citet{AMIBA}), \citet{10m}, \citet{Planck}, etc. and the
current CBI (\citet{CBI}) are
capable of detecting this angular scale.  Using multi-frequency
information, the SZ fluctuation can be disentangled from the primary
anisotropies at all angular scales \citep{Cooray00a}.

On the theory part, various approaches have been explored to compute
the SZ angular power spectrum.  Amongst these, the Press-Schechter
formalism is perhaps the most widely used due to its versatility and
ease of implementation \citep{Cole88,Makino93,Atrio99,Komatsu99,Cooray00b}. The authors above adopted the cluster gas model,
while \citet{Silva99,Refregier99,Seljak00} used simulations and
\citet{Cooray00a} used a simplified gas pressure bias model motivated
from simulations to probe the statistics of the SZ effect.

One of the shortcomings of the SZ effect is its lack of redshift
information.  Since scattering is independent of redshift, the SZ
effect on one hand allows a direct probe of the IGM to high redshift,
on the other hand makes it challenging to disentangle the
contributions arising from different redshifts.  In this paper, we
apply the hierarchical clustering model and non-linear perturbation
theory to directly compute the SZ effect.  This analytical method
enables us to to extract distance information of intergalactic gas
from the cross correlation of the SZ effect with galaxy surveys and at
the same time, check the results of both simulations and the
Press-Schechter formalism. In section   
\ref{sec:ps}, we develop our gas pressure model and in section
\ref{sec:stat} the SZ statistics including power spectrum and
bispectrum.  Section \ref{sec:cross} contains our results on the
SZ-galaxy cross correlation and the method to extract the redshift
distribution of the SZ effect.  The effect of non-gravitational
heating and the method to extract it from overall gas pressure
power spectrum are discussed in section \ref{sec:NGheating}. We
discuss the potential
inaccuracies arising from the approximations in section
\ref{sec:discussion}.  The paper concludes with section \ref{sec:conclusion}.

\section{Gas pressure power spectrum}
\label{sec:ps}

  The temperature distortion caused by the SZ effect \citep{Zeldovich69}
is:
\begin{equation}
\Theta(\hat{n})\equiv\frac{\Delta T_{CMB}(\hat{n})}{T_{CMB}}=-y(\hat{n})
\frac{xe^x}{e^{x}-1}\left[4-x/\tanh(x/2)\right] \equiv -2yS(x)
\end{equation}
where $\hat{n}$ is the direction on the
sky, $x=h\nu/(kT_{CMB})$, and the scattering function
$S(x)\rightarrow 1$ when
$x \ll 1$ (Rayleigh-Jeans tail). In this limit the SZ effect
results in an apparent
cooling of the CMB background. The ``$y$'' parameter is defined as
\begin{equation}
y(\hat{n})=\frac{\sigma_T}{m_e c^2} \int_0^{l(z_{cmb})} n_e kT_g dl=
\frac{\sigma_T}{m_e c^2} \int P_e(\hat{n}) dl=\frac{\sigma_T}{m_e c^2} \int
\bar{P_e} y_p(\hat{n}) dl
\end{equation}
Here, $T_g$ and $n_e$ are the temperature and number density of free electrons,
respectively.  $P_e$ is the gas
pressure. $y_p=P_e/\bar{P_e}=(1+\delta_g) T_g/\bar{T_g}$  and
$\bar{T_g}\equiv\langle(1+\delta_g)T_g\rangle$ is the gas density weighted mean
temperature. $dl=a(z) C(x) dx$ is the proper distance, $a(z)$ is the
scale factor, $x(z)$ is the
comoving distance, 
\begin{equation}
\label{eqn:cx}
C(x)=\frac{1}{\left[1-K(x/R_0)^2\right]^{1/2}}
\end{equation}
describes 
the geometric effect of the curved universe, $K=-1,0,1$ for open, flat
and closed universes respectively,
and  $R_0=\frac{c}{H_0} (1-\Omega_0)^{-1/2}$ is the curvature radius.
$\rho_0=\rho_c \Omega_0$ is the present cosmological matter density.

\citet{Pen99} showed that the intergalactic medium has most likely
been preheated by non-gravitational sources to $\sim 1$ keV per
nucleon in order  to be consistent  
with the observed upper bounds from the X-ray background.
We adopt the model of \citep{Pen99} to express the gas distribution as a
convolution of the matter distribution:
\begin{equation}
\label{eqn:gasw}
\delta_g(x)=\int \delta(x')W_g(|x-x'|)d^3 x'
\end{equation}  
This equation expresses gas as being less clumped than
dark matter, partly due to the required preheating. The effective
radius in the top-hat window function, which is the gas 
heating radius, has the typical value  $\sim 1 h^{-1}$ Mpc
from the X-ray background constraint \citep{Pen99}. For simplicity, we adopt
the Gaussian window $W_g(r)=
\exp(-r^2/2r_g^2)/(\sqrt{2\pi}r_g)^3$ ($r_g 
\sim 1/3 h^{-1}$Mpc corresponds to $1h^{-1}$ Mpc top hat 
window).  The window function for 
specific heating models can be obtained from hydrodynamic 
simulation  by comparing the gas power
spectrum to the dark matter power spectrum  (Ma and Pen, 2000).
We also need to relate the gas temperature with the density. First, we
consider the gravitational heating.   
We adopt the cosmic energy theorem \citep{Peebles80} for the gas
temperature model, which specifies the ratio between gravitational
binding energy and total kinetic energy $K$.  The pressure depends on the
thermalized fraction of $K$.  The translational kinetic energy
is thermalized from the
energy released when particles shell cross.  A model of the
thermalized energy is thus given by the difference in energy between 
two particles separated by a non-linear scale in Lagrangian space,
which is the distance at which they can be expected to have shell
crossed.  The exact procedure amounts to solving the non-linear
evolution equations directly.  But we can treat the effect
statistically in a linear fashion.  In the initial linear evolution, the
gravitational potential remains constant.  After virialization, the
gravitational energy at a fixed location remains almost
constant.  In an Eulerian description, we can describe the energy of
particles at a final virialized location as the energy released as a
particle travels from its initial position to the final virialized
location.  While the initial position is not exactly known, we take a
spherical average over the non-linear scale to average over all
possible initial locations.  We thus have
\begin{equation}
kT_g=\frac{1}{6} \frac{4 m_H}{3+5 X} (\Psi-\bar{\Psi})
\label{eqn:temp}
\end{equation}
$\Psi$ is the gravitational potential, 
$\bigtriangledown^2 \Psi=-4\pi G \bar{\rho} a^2
\delta$. $\bar{\Psi}(x)=\int \Psi(r)W_e(|x-r|)d^3r$ (hereafter call
the ``electron window function'') is the potential
averaged over the non-linear scale $r_e$. We choose
a Gaussian window function $W_e(r)$ with the the non-linear scale
$r_e$, which for $z=0$ is $r_e \sim 5 h^{-1}$Mpc. Hereafter we will
adopt this value of $r_e$. $X=0.76$ is the mass fraction of the
hydrogen in baryonic matter. Then, $(kT_g)_k \propto \delta(k)
\left[1-W_e(k)\right]/k^2 \equiv \delta(k) f_e(k)$. Here, $W_e(k,z)=\exp(-k^2
r_e^2(z)/2)$ is the Fourier transform of the electron window function.  
Equation (\ref{eqn:temp}) has some unphysical statistical properties.
The spatial average of the temperature, for example, is exactly zero,
and for our model using Gaussian random fields, it will be negative in
half the volume.  But for purposes of modeling the SZ effect,
temperature is only observable when multiplied by density, and regions
of positive temperature will have high density, while the negative
temperature regions only contribute negligibly to the SZ effect.  This
model is not meant to be an exact description, but hopefully captures
the statistical properties, while being simple and thus exactly solvable.

\begin{figure}[ht]
\figurenum{1}
\plotone{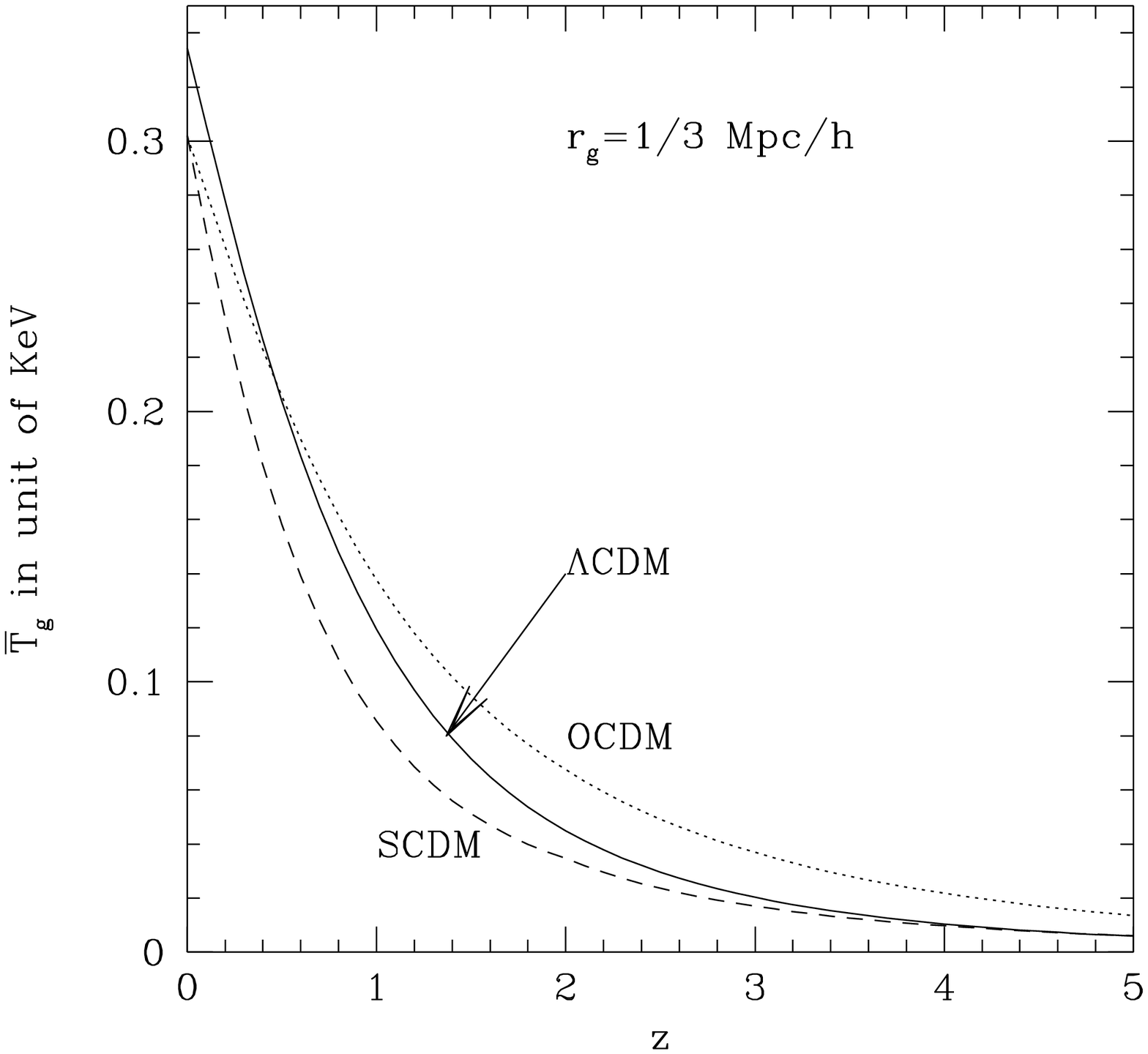}
\caption{The evolution of the gas density averaged
temperature. Smaller $r_g$ produces
higher $\bar{T_g}$. \label{fig:temp}}
\end{figure}

In order to estimate the accuracy of these assumptions, we first
compute the gas 
density weighted temperature and the mean $y$ parameter. We define
$\delta(x)\equiv (2 \pi)^{-3} \int\delta_k \exp(-i k\cdot x)d^3 k$, the
power spectrum $P(k)\equiv\langle|\delta_k|^2\rangle$ and the variance
$\Delta^2(k)\equiv k^3 P(k)/2 \pi^2$.  We adopt the initial
power spectrum \citep{Peebles83, Davis85} $P_{linear}(k)\propto 
k^{1+\alpha}/(1+1.7 k+9 k^{1.5}+k^2)$ (here, $k$ is in unit of
$\Omega_0 h^2$/Mpc and we choose the Harrison-Zel'dovich-Peebles scale
invariant spectrum $n=-1$ corresponding to $\alpha=0$), 
cluster-normalized density fluctuation 
at $8 h^{-1}$Mpc by $\sigma_8=0.53 \Omega^{-\beta}$ [$\beta=0.53$ for
$\Lambda$CDM and $\beta=0.45$ for OCDM ] \citep{Pen98}  and
the \citet{Peacock96} fitting formula to convert the
linear $P_{linear}$ to the  nonlinear $P(k)$. Then 
\begin{equation}
\bar{T_g}=\langle(1+\delta_g)T_g\rangle=4 \pi G \rho_0 \frac{1}{6} \frac{4
m_H}{3+5X} (1+z)\int_0^{\infty} \Delta^2(k) W_g(k) f_e(k) \frac{dk}{k}
\end{equation}
\begin{equation}
\bar{y}= \frac{\sigma_T}{m_e c^2} \int \bar{P_e}
dl=\frac{\sigma_T}{m_e c^2} f_g \frac{\rho_B(0)}{2 m_H/(1+X)}
\int_0^{x(z_{cmb})} 
\bar{T_g}(z) (1+z)^2 C(x)dx
\end{equation}
$W_g(k)$ is the Fourier transform of the gas window function
in equation (\ref{eqn:gasw}), 
$\rho_B=\rho_c \Omega_B$ is the present baryonic matter density and $f_g$
is the mass fraction of baryonic matter in gaseous   
form. Since cluster gas fractions \citep{Danos98} are comparable to
the baryon fraction obtained from big bang nucleosynthesis, we expect
that 
$f_g \sim1$.  We use $f_g=0.9$, $\Omega_B h^2=0.02$ and the
dimensionless Hubble constant $h=0.67$. The results for
SCDM, OCDM and  $\Lambda$CDM are shown in table \ref{tble:tbar} and
figure \ref{fig:temp}.

 In the absence of non-gravitational heating effects,
we find the present $\bar{T_g}$ is around $0.35$ keV  and the mean
SZ temperature distortion $\bar{\Theta} \sim 5\times 10^{-6}$K . They are all 
consistent with the simulations and Press-Schechter formalism
results \citep{Refregier99,Seljak00}. 

\begin{table}
\begin{tabular}{|c|c|c|c|c|c|c|}  \tableline
Cosmologies
&\multicolumn{2}{c|}{SCDM}&\multicolumn{2}{c|}{OCDM}&\multicolumn{2}{c|}{$\Lambda$CDM}
\\ \tableline
$\Omega_0$ &\multicolumn{2}{c|}{1} &\multicolumn{2}{c|}{0.37}
&\multicolumn{2}{c|}{0.37} \\ \tableline
$\sigma_8$ &\multicolumn{2}{c|}{0.53} &\multicolumn{2}{c|}{0.83}
&\multicolumn{2}{c|}{0.90} \\  \tableline
$r_g/(h^{-1}$Mpc) &$1/3$&$1/6$&$1/3$ &$1/6$&$1/3$&$1/6$\\ \tableline 
$\bar{T_g}(z=0)$/keV  &0.3   &0.35    &0.30  &0.35  &0.33  &0.38
\\ \tableline 
$\bar{y}\times 10^6$  &1.4 & 1.7  &2.9   &3.5  &2.3   &2.6       \\
\tableline
\end{tabular}
\caption{Present day mass weighted average gas temperature
$\bar{T_g}(z=0)$, $\bar{y}$,
and temperature distortion skewness parameter $\Theta_3\equiv
\frac{\langle(\Theta-\bar{\Theta})^3\rangle}
{\langle(\Theta-\bar{\Theta})^2 \rangle^2}$ for various cosmologies.
Two different gas-mass relations are used ($r_g=1/3,1/6$), which only
has a small effect on the results.} 
\label{tble:tbar}
\end{table}

We define the pressure correlation function
$\xi_p(r)\equiv
\langle y_p(x)y_p(x+r) \rangle \equiv\langle [1+\delta_g(x)]
T(x)[1+\delta_g(x+r)] 
T(x+r)\rangle/\bar{T_g}^2$. \footnote{Our definition is different to
the usual definition of $\xi_p(r)\equiv
\langle \delta_p(x)\delta_p(x+r) \rangle$. Here, $\delta_p\equiv
y_p-1\equiv (P_e-\bar{P_e})/\bar{P_e}$. These
two definitions only differ by a constant $1$ and the resulting power
spectrums only differ when $k=0$ by a Dirac function. We choose this
definition because this is the most convenient way to deal with SZ
effect.} Then the pressure power spectrum   
\begin{eqnarray}
\label{pressure}
P_p(k,z)&=&\left[\int \Delta^2(k,z) W_g(k) f_e(k,z) \frac{dk}{k}\right]^{-2}
\times \nonumber \\ 
&&\left( \frac{1}{(2 \pi)^6} \int
B_4({\bf k_1},{\bf k_2},{\bf k_3},{\bf k_4};z)  W_g(k_1) f_e(k_2,z) W_g(k_3)
f_e(k_4,z) d^3 k_2 d^3 k_4 \right. \\
 &&+\frac{1}{(2 \pi)^3} \int W_g(k_1) P(k_1,z) f_e(k_2,z)
P(k_2,z)\left[W_g(k_1)f_e(k_2,z)+W_g(k_2)f_e(k_1,z)\right]d^3 k_2 \nonumber\\
 &&+2 \frac{1}{(2 \pi)^3} \int B_3(k_1,k_2,k;z) W_g(k_1) f_e(k_2,z) f_e(k,z)
d^3 k_2 \nonumber \\
&&+ P(k,z) f_e^2(k,z)\ ) \nonumber
\end{eqnarray}
Statistical isotropy implies, $\bf{k_1}+\bf{k_2}=-k_3-k_4=k$. The
first term in the parentheses is from the 
non-Gaussian (connected) part of the four-point correlation
$\langle\delta_g(x)T(x)\delta_g(x+r)T(x+r)\rangle$. $B_4$ is the
density polyspectrum. The second term is from the Gaussian part of
$\langle\delta_g(x)T(x)\delta_g(x+r)T(x+r)\rangle$. This part also
produces a term $(2 \pi)^3 \delta_D(\bf k)$ in $P_p(k)$, which only
affects the spatially averaged quantities such as $\langle P_e^2
\rangle$ and  
will be included explicitly only when it is observable.  The
third term is from the three-point correlation
$\langle\delta_g(x)T(x)T(x+r)\rangle+\rm permutation$.  $B_3$ is the
density bispectrum. The last term is from the raw temperature correlation
$\langle T(x)T(x+r)\rangle$.

We adopt the hierarchical model \citep{Fry84} which allows all higher
order correlations to be expressed as the sum of products of two point
correlations over all configurations, so
\begin{equation}
B_N({\bf k_1,...,k_N})=\sum_{a=1}^{t_N}  Q_{N,a}
\sum_{{\rm labelings}} \prod_{{\rm edges}}^{N-1} P_{AB}
\end{equation}
$P_{AB}$ is
the two point power spectrum. $a$ denotes different
configurations. Coefficients $Q_{N,a}$ generally depend on 
configurations. In the highly non-linear regime, they
degenerate \citep{Scoccimarro99}. Because the SZ effect is mostly
contributed by the  highly nonlinear  regime, we adopt the saturated
$Q_{AB}$ in the highly non-linear regime:  
$Q_4^{sat}=\frac{27(1-2^{n-1})+3^n+\frac{1}{2} \times 6^n}{1+6\times
2^n+3^{n+1}+6^{n+1}}$ and $Q_3^{sat}=\frac{4-2^n}{1+2^{n+1}}$ from
HEPT (the Hyper-Extended Perturbation Theory) \citep{Scoccimarro99}.  Here,
$n$ is the linear power spectrum 
index and we choose the value of $n$ at $k=\sum_{i=1}^N (k_i/N)$.  In this
framework,  
\begin{eqnarray}
B_3(k_1,k_2,k_3)&=&Q_3 (P_1P_2+P_2P_3+P_3P_1)\nonumber \\
B_4(k_1,k_2,k_3,k_4)&=&Q_4\left[P_1 P_2 P_3+(3 {\rm \ Permutations})+P_1
P_{12} P_4+(11 {\rm \ permutations})\right]
\end{eqnarray}
where $P_i=P(k_i)$ and $P_{ij}=P({\bf|k_i+k_j|})$.

\begin{figure}
\figurenum{2}
\plotone{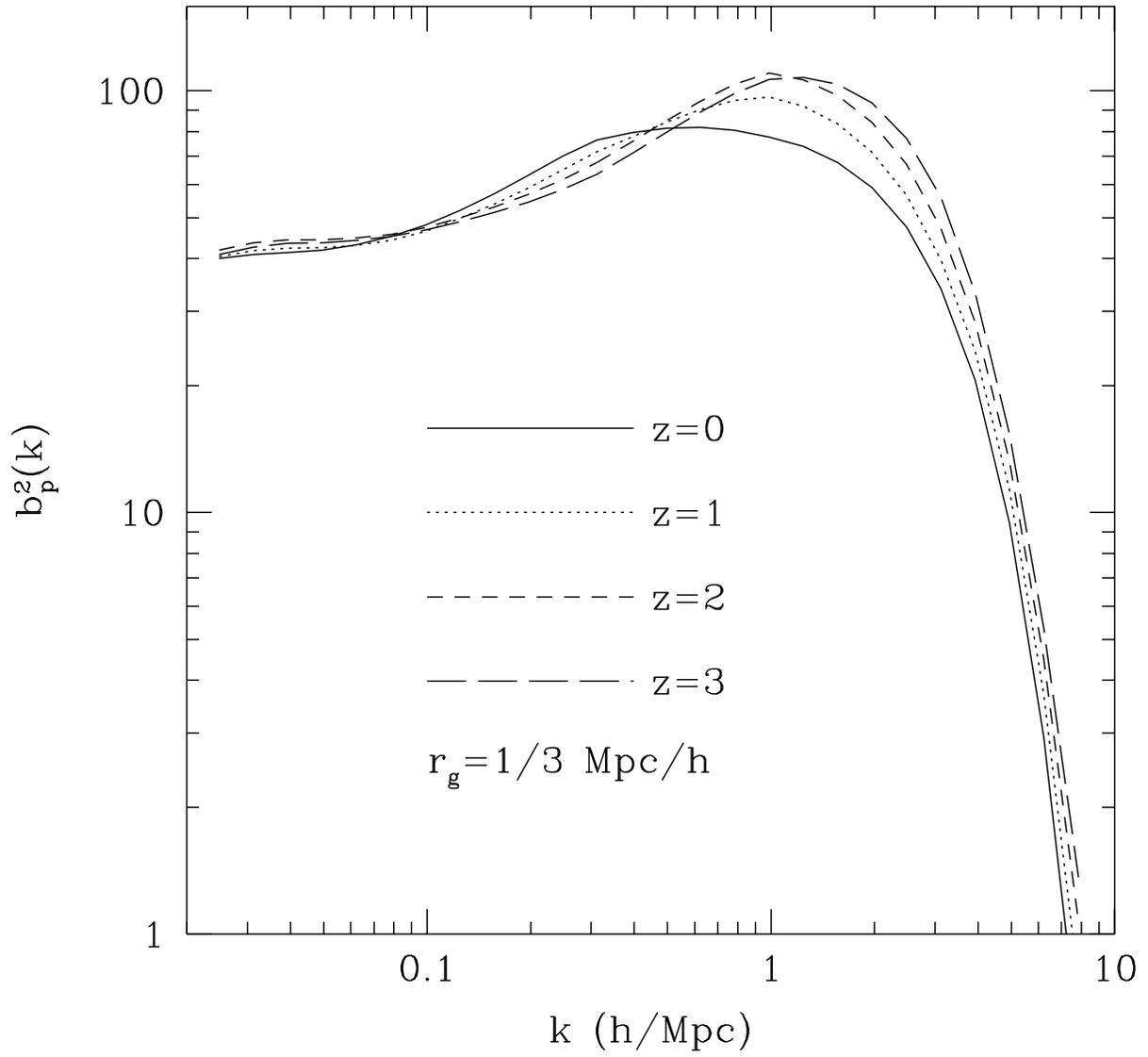}  
\caption{Gas pressure bias in $\Lambda$CDM model. Results for OCDM and
SCDM are similiar. \label{fig:bias}}   
\end{figure}

Results are shown in figures \ref{fig:bias} and
\ref{fig:pressure}. Calculations show that the non-Gaussian term is  
dominant in the spectrum even when 
$k$ is small and this non-Gaussian term itself is mostly contributed 
by the nonlinear regime. Besides the reason that the electron window
function sweeps off most contribution from the linear regime, these behaviors
reflect the  strong correlation between $\delta$ and $T$ and
the domination of the highly nonlinear (non-Gaussian)
regime.  This argues that the isotropic simplification for $Q_N$ is at
least self-consistent.   We define the gas pressure bias $b_p$  as
$b_p^2(k,z)\equiv P_p(k,z)/P(k,z)$. The amplitude of  $b_p$
varies only weakly with time. This is an expected consequence of the 
hierarchical model, argued as followed. The time dependence through 
the density power spectrum is basically canceled because Eq. (\ref
{pressure}) shows that the dependence of $P_p(k,z)$ over time is
roughly $P_p(z)\propto B_4(z)/P^2(z)$. The hierarchical model implies
that roughly $B_4(z) \propto P^3(z)$. Then $b^2_p(z)\equiv
P_p(z)/P(z) \propto P^0(z)$. The only remaining time 
dependence arises from the evolution 
of the electron window function. Its effect is to eliminate the
contribution from scales larger than several $r_e$. Since  $r_e$ decreases with
redshift, its evolution moves the peak of the bias to
smaller scales, which should have higher bias. The bias
approaches a constant ($\sim 6$ for all three cosmological models) at
large scale, but it drops quickly at very small scale (due to the gas
window function). This result is consistent with
\citet{Refregier99}. One point worthy of mention is that, the behavior
that $b_p$ approach a constant at large scale does not validate
the constant bias model, which always produces a unit pressure-density cross
correlation coefficient. But as we will see in Sec. \ref{sec:cross}, this
is not the case. The pressure fluctuation
variance peaks roughly 
at the scale of cluster core radii ($\sim 0.13 h^{-1}$ Mpc), which is
the direct result of the gas window function ($r_g \sim 1/3 h^{-1}$ Mpc).

\section{Statistics of Sunyaev-Zel'dovich effect on CMB anisotropies}
\label{sec:stat}

The SZ effect on CMB anisotropies is fully described by the $n$-point
correlation functions between temperature distortion $\Theta$ in
different sky directions. The CMB anisotropy is decomposed into the
harmonic components by:  
\begin{equation}
\Theta(\hat{q})\equiv\frac{\delta T}{T}(\hat{q})=\sum^{l,m}
a_{lm}Y_{lm}(\hat{q}) 
\end{equation} 
The CMB power spectrum is defined as $C_l\equiv\langle \sum_{m=-l}^l a_{lm}
a_{lm}^*\rangle/(2l+1)$.  The bispectrum is
defined as $B_{l_1 l_2 
l_2}^{m_1m_2m_3}\equiv \langle a_{l_1}^{m_1}a_{l_2}^{ m_2} a_{l_3}^
{ m_3}\rangle$. They  have been discussed by many authors as mentioned in
the introduction section.  Our model successfully reproduces these
results. 
In contrast to the Press-Schechter models,
our model relies only on direct statistical correlators that have been
proposed semi-analytically with only a few free dimensionless
parameter, the $Q$ saturation values, that have been verified in simulations.  In the
Press-Schechter model, the shape of the correlation function depends
directly on the assumed structure of halos, which is a free function
of the model and empirically measured in simulations.  The radially
averaged structure of halos, however, is not necessarily a direct
predictor of correlations, unless one assumes halos to have no
substructure. It is reassuring to see that our approach, which
originates from the diametrically opposite
theoretical bases from Press-Schechter formalism, produces consistent
results and agrees with simulations. 
\subsection{SZ power spectrum}
\begin{figure}
\figurenum{4}
\plotone{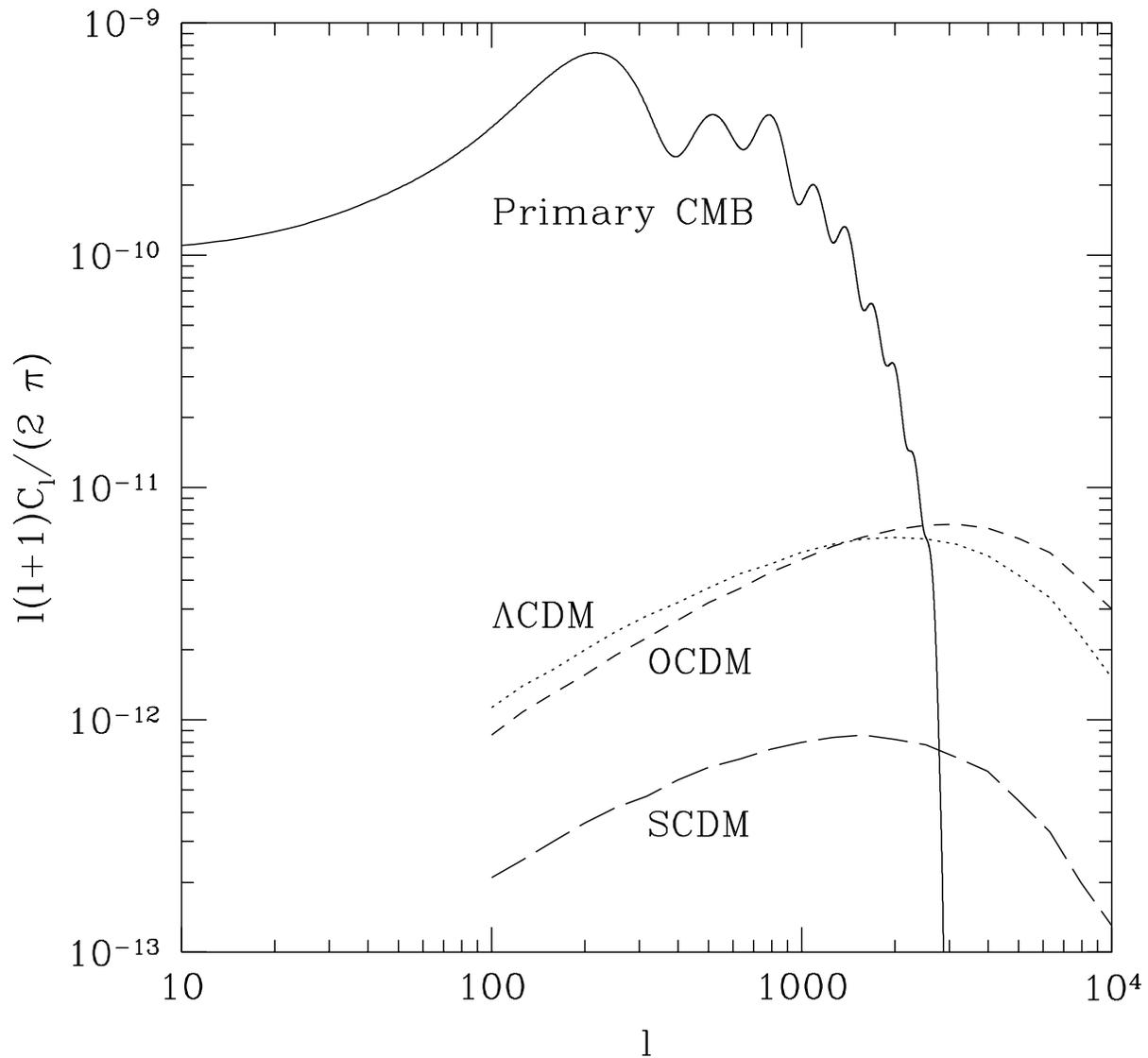}
\caption{CMB anisotropy caused by SZ effect in various
models. $r_g=1/3 h^{-1}$Mpc. Smaller $r_g$ will move
peaks to larger $l$ ( smaller angle) and larger amplitude. \label{fig:cl}}     
\end{figure}

\begin{figure}
\figurenum{5}
\plotone{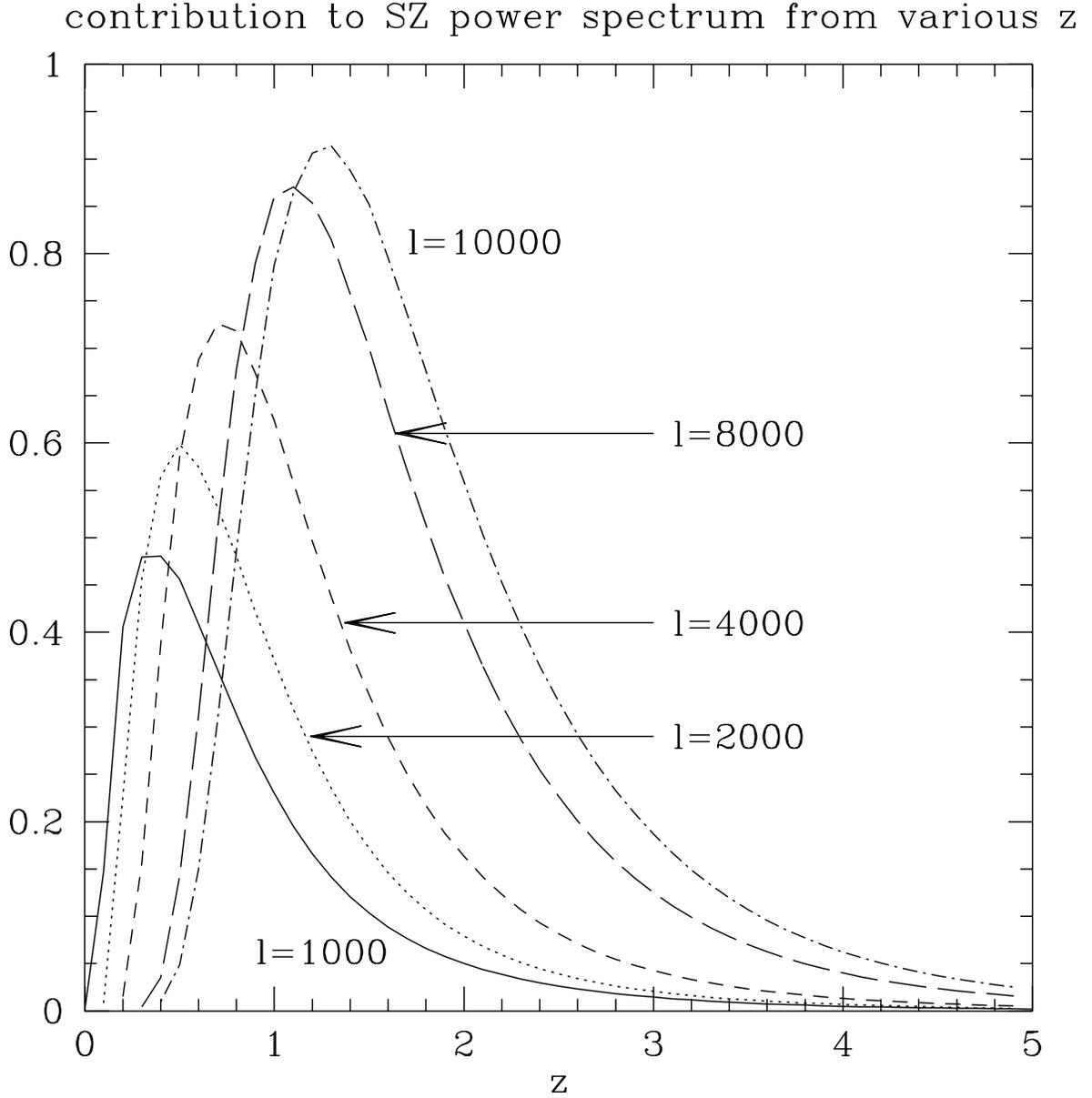}
\caption{SZ effect contribution from 
different redshift, defined as $\frac{\Delta^2_y(k,z)|_{k=l/x} C(x)
x(z) \frac{dx}{dz} z}{\int_0^{z_{cmb}} \Delta^2_y(k,z)|_{k=l/x} C(x)
x(z) \frac{dx}{dz}dz}$. We show a $\Lambda$CDM model
($r_g=\frac{1}{3}/h$ Mpc). Results  for SCDM and OCDM are 
similiar.  Smaller $r_g$ moves peaks to smaller $z$ and lower
height. \label{fig:zcon}} 
\end{figure}

Since we are interested in small angular scale where the
line-of-sight CMB survey
depth is much  larger  than any correlation length, we adopt  Limber's
equation \citep{Peacock99}. In the Rayleigh-Jeans  
region we get the SZ power spectrum:
\begin{equation}
\label{eqn:cl}
C_l=\frac{64\pi^2}{(2l+1)^3}\int_0^{x(z_{cmb})}\Delta^2_{SZ}(k,z)|_{k=l/x}C(x)
x(z)dx(z) 
\end{equation}
where
\begin{equation}
\Delta_{SZ}^2(k,z)\equiv \left(\frac{\bar{P}_e(z) \sigma_T }{m_e c^2
(1+z)}\right)^2
\frac{1}{2 \pi^2} k^3 P_p(k,z)
\end{equation} 
The results are shown in figure \ref{fig:cl} and \ref{fig:zcon}. We find that (1) The SZ effect exceeds the 
primary CMB anisotropy at $l\sim 2000$. (2) The $C_l$ of SCDM is much
smaller than that of OCDM and $\Lambda$CDM. This is due to the smaller
$\sigma_8$ and faster drop of the gas temperature in SCDM with
increasing redshift. (3) Because
$P(k)\propto \sigma_8^{2--3}$  ($2$ in linear regime and $3$ in stable
clustering  regime) and $P_p(k)\propto B_4 \propto P^3(k)$, we at once
find that $C_l\propto \sigma_8^{6--9}$. (4) The peak contribution to
$C_l$ depends on the angular scale, 
varying from $z\sim 0.4$  at $l\sim1000$ to $z\sim 1.5$ for $l\sim
10000$. Smaller angles probe the more distant 
universe.  These results are consistent with existing simulations and
Press-Schechter formalism results.  
\subsection{SZ bispectrum}
The SZ bispectrum is the projection of the pressure bispectrum
$B^p_3(k_1,k_2,k_3)$ defined by 
\begin{equation}
\langle y_p({\bf k_1}) y_p({\bf k_2}) y_p({\bf k_3})
\rangle\equiv B^p_3({\bf k_1,k_2,k_3}) \delta_D({\bf
k_1+k_2+k_3})
\end{equation}  Since $y_p\propto \delta^2$, $B^p_3$ is related to
6-point density correlation  function and in principle can be computed
using our model. But due to the complexity of the 6-point hierarchy and
the absence of extensive numerical tests, we adopt the approach of
\citet{Cooray00a} and  
assume a simple pressure bias model where $\delta_p(k,z)=b_p (k,z)
\delta_k(z)$ with $b_p^2(k,z)\equiv P_p(k,z)/ P(k,z)$.  The pressure
bias model states that the pressure is a linear convolution over the
density.  We note that this model builds in certain assumptions, which
are invalid for treating cross-correlation issues address in
section \ref{sec:cross}.  It presumably provides a quick order of
magnitude estimator, but fails on qualitative issues like
galaxy-SZ cross-correlation coefficients.

In the bias model the pressure bispectrum is given as
\begin{equation}
B_3^p({\bf k_1,k_2,k_3};z)=\prod^3 b_p(k_i,z)B_3({\bf k_1,k_2,k_3};z).
\end{equation}
The SZ skewness parameter,
\begin{equation}
\Theta_3\equiv
\frac{\langle(\Theta-\bar{\Theta})^3\rangle}{\langle(\Theta-\bar{\Theta})^2\rangle^2} 
\label{eqn:t3}
\end{equation}
is the easiest observable projection of the
SZ bispectrum, so we show its computation  as an example.
 
Instead of working in multipole space, we derive following equations
adopting the same approximation as Limber's equation:
\begin{equation}
\langle\Theta(\hat{n})^2\rangle=\bar{\Theta}^2+\frac{4}{2
\pi} \int 
\left(\frac{\sigma_T \bar{P_e}(z)}{m_ec^2 (1+z)}\right)^2 C(x) dx(z)
\times \left[\int b_p^2(k,z) P(k,z) k dk\right]
\end{equation}
\begin{eqnarray}
\langle\Theta(\hat{n})^3\rangle&=&\bar{\Theta}^3-\frac{8}{(2 \pi)^2} \int
(\frac{\sigma_T \bar{P_e}(z)}{m_ec^2 (1+z)})^3 C(x) dx(z) \nonumber \\
&\times&
\left[\int \prod_{i=1}^{3}b_p(k_i,z) B_3({\bf k_1,k_2,k_3};z) k_1 dk_1 k_2
dk_2\right]
\end{eqnarray}
where $\bf k_3=k_1+k_2$ and $\bf k_1 \parallel k_2$. $\bar{\Theta}=-2
\bar{y} \sim 5 \times 10^{-6}$ is the mean temperature SZ distortion.
$\bar{\Theta}^2$ and $\bar{\Theta}^3$ are from the corresponding
Gaussian  term in the correlation function, as explained in
Eq. (\ref{pressure}). As a reminder, $C(x)$ is
the geometric function defined in Eq. (\ref{eqn:cx}). 
Though the unsmoothed $\Theta_3$ is not directly observable
due to the absence of the window function, from the theory viewpoint,
it represents the raw non-gaussianity properties.  Our calculations find
that $\Theta_3 \sim 10^{6}$ (Table \ref{tble:tbar}).
We defined the skewness parameter (\ref{eqn:t3}) as the 
ratio of differential measurements, which might result from
interferometers in the Raleigh-Jeans regime.
For comparison, \citet{Cooray00a} and \citet{Cooray00b} considered the absolute
temperature distortion and defined the skewness
parameter as $\langle\Theta^3\rangle/\langle
\Theta^2\rangle^2$. 
With multi-frequency information the absolute $y$ could be
measured at each point in space,
and that definition of skewness could in principle be observed.  
The results are not too different: In our
model,  $\langle\Theta^3\rangle/\langle 
\Theta^2\rangle^2 \sim 10^5$, which is similar to the result of
\citep{Cooray00b} with the maximum virialized mass $\sim
10^{15}M_{\odot}$. Further work considering window function filtering and the
general calculation of the SZ bispectrum can be done following the method
of \citet{Cooray00a}.   

\section{Extracting redshift information from SZ-galaxy cross correlation}
\label{sec:cross}

   In the SZ effect, the redshift distribution of intergalactic gas is
lost. Taking advantage of the cross correlations with other surveys
having redshift information, we may be able to statistically extract
IGM 3-D correlations and  evolution. Because the SZ effect is mostly
contributed by nonlinear structures at $z\sim 0.5-2$, it should have a
strong cross correlation with galaxies at that redshift range.  The general
idea is to use a galaxy survey with coarse redshift information, where
the distance information only needs enough accuracy to resolve the
time evolution of the correlation function, for example $\Delta z \sim
0.2$, which can be reached by photometric redshift surveys.  One then
correlates each redshift bin with the SZ map, and obtains the relative
contribution of that SZ slice to the total projected map.

Several questions that pose themselves in this procedure are: How big
a survey area and depth does one need?  What fraction of the total SZ
fluctuations can be accounted using cross-correlations?  Are there
optimal procedures of cross-correlating? Since SZ fluctuations are a
non-linear function of the galaxy field, what fraction of the signal
will ever be accounted for by this approach?  Do we expect the SZ-galaxy
cross-correlation coefficient to depend on redshift?  How big an
uncertainty might arise from a time and scale dependent galaxy bias?
In this section we will address each of these questions using our
model, and obtain quantitative estimates.

\begin{figure}
\figurenum{3}
\plotone{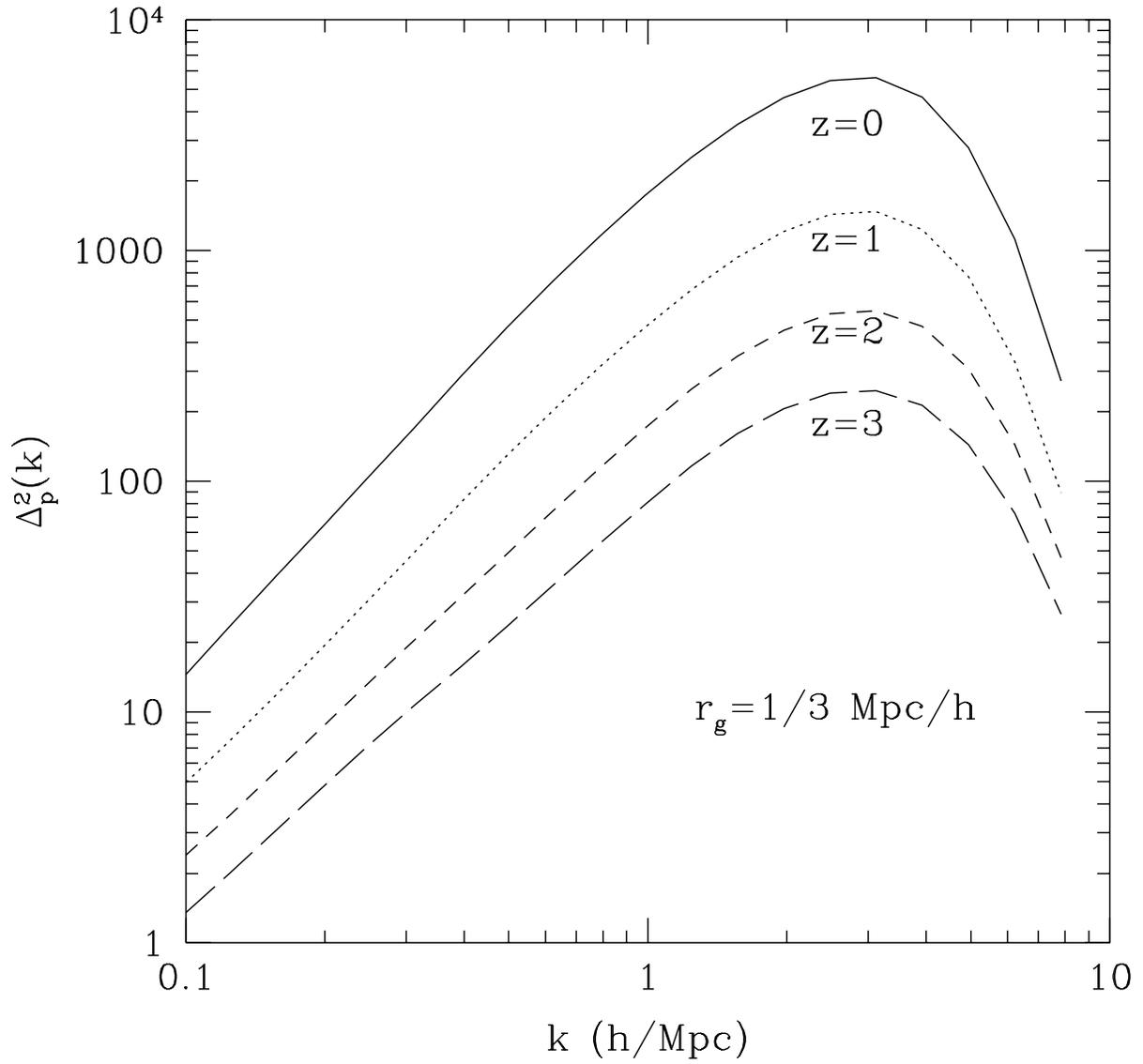}
\caption{Gas pressure variance $\Delta_p^2\equiv\frac{k^3}{2 \pi^2}
P_p(k)$. Results for OCDM and SCDM are
similiar. \label{fig:pressure}} 
\end{figure}

\begin{figure}
\figurenum{6}
\plotone{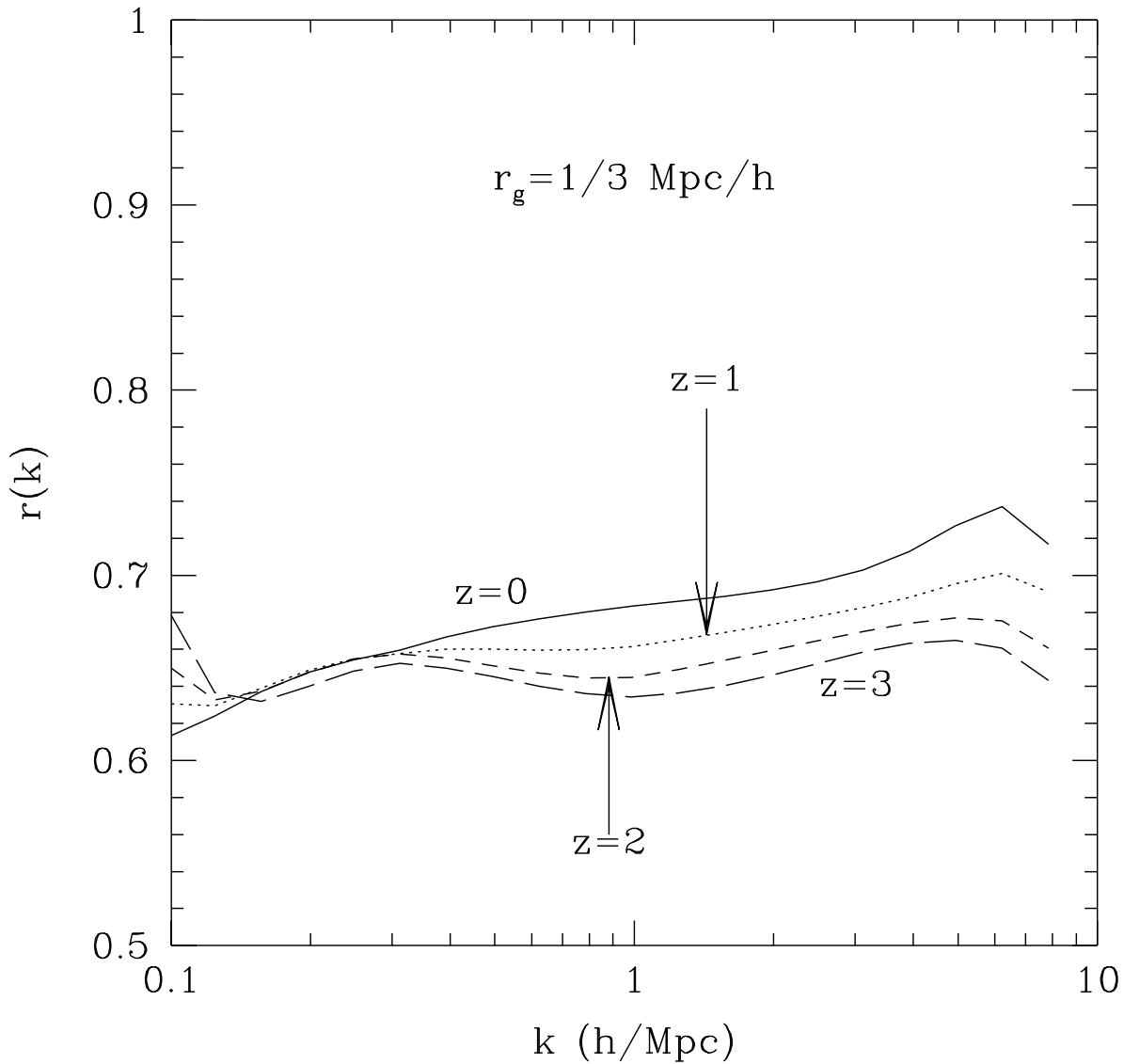}
\caption{SZ-Galaxy cross correlation coefficients in real space
for a $\Lambda$CDM  model. Results have almost
no dependence on cosmological models and gas model
parameters. The region shown is the most relevant region of SZ effect,
as shown in figure \ref{fig:pressure}. The behavior of $r(k,z)$ in
other regions is estimated in Sec. \ref{sec:cross}.  \label{fig:r}}  
\end{figure}

We assume a linear bias model for the  galaxy number overdensity
$\delta_G=b \delta$. The bias $b$ is
taken to be a constant for simplicity. Then, the projected galaxies number
overdensity can be expressed as 
\begin{equation}
\frac{\Delta n}{n}(\hat{q})=\int_0^{x(z_G)} \delta_G(x\hat{q}) \phi(z)
C(x)dx(z).
\end{equation}
$\phi(z)$ is the selection function (which we will want to
vary {\it a posteriori}  using the photometric redshift information),
$z_G$ is the redshift
survey depth and $C(x)$ is the geometric function defined before.
  The corresponding multipole moments are:
\begin{equation}
\label{eqn:galaxy}
C_l^G=\frac{16 \pi^2}{(2l+1)^3}\int_0^{x(z_G)} \Delta^2_G(k,z)|_{k=l/x}
\phi^2(z) C(z) x(z)dx(z)
\end{equation}
where $\Delta^2_G=b^2 \Delta^2$. The multipole moments of SZ-galaxy
cross correlation are: 
\begin{equation}
\label{eqn:szg}
C_l^{SZ,G}=\frac{2 \times 16 \pi^2}{(2l+1)^3}\int_0^{x(z_G)}
\Delta^2_{SZ,G}(k,z)|_{k=l/x} \phi(z) C(z) x(z)dx(z)
\end{equation}
with $\Delta^2_{SZ,G}(k,z)=\frac{\bar{P_e} \sigma_T}{m_ec^2 (1+z)} \frac{1}{2 
\pi^2} k^3 P_{p,G}(k)$. $P_{p,G}$ is the Fourier transform of $\langle y_p({\bf x})\delta_G({\bf x+r})\rangle$.   
\begin{equation}
P_{p,G}(k,z)=b\times  \frac{P(k) f_e(k) +\frac{1}{(2 \pi)^3} \int
B_3(k_1,k_2,k) 
W_g(k_1)f_e(k_2) d^3 k_2}{\int \Delta^2(k) W_g(k)
f_e(k)\frac{dk}{k}}
\end{equation}
We have used $\bf k=k_1+k_2$.
It is useful to define the cross correlation coefficient 
\begin{equation}
r(k,z)\equiv
\frac{\Delta^2_{SZ,G}(k,z)}{\Delta_{SZ}(k,z)\Delta_G(k,z)} \equiv \frac{P_{p,G}(k,z)}{\sqrt{P_p(k,z)P(k,z)}}
\label{eqn:rkz}
\end{equation}
In $P_{p,G}$, the $B_3$ term is dominant.  In $B_4$ there are 16
hierarchical terms and in $B_3$ there are three 
hierarchical  terms. Different term dominates in different
regions. Calculation shows that, roughly
there are three regions: (a) $k\lesssim 0.1 {\rm h/Mpc}$. 4 terms
( $P_{12}(2 P_1P_4+P_1P_3+P_2P_4)$ ) dominate $B_4$ and 2 terms
( $P_3 P_1+P_3 P_2$ )   dominate $B_3$. $r\simeq
Q_3/Q_4^{1/2} \simeq 0.9$ ($n\sim -1.5$)\footnote{Though $k$
is small, contribution to $P_p(k,z)$ and $P_{p,G}(k,z)$ are mostly
from nonlinear regimes with $-2 \lesssim n \lesssim -1$.}.
(b) $k\lesssim  0.1 {\rm h/Mpc} \lesssim 10 {\rm h/Mpc}$.  Each 
hierarchical term  in $B_3$ and $B_4$ has about the same contribution
to $P_{p,G}$ and $P_p$, respectively. $r(k,z)\sim 3/4 \times
Q_3/Q_4^{1/2}\sim 0.7$  ($n\sim-1.5$). This region contributes
most of the SZ effect, as seen from figure \ref{fig:pressure}. Unless
explicitly notified, hereafter we will adopt the 
value of $r$ in this region.  We only show the result
of this region in Figure \ref{fig:r}.
(c) $k \gtrsim 10 {\rm
h/Mpc}$. This is the opposite case to the case (a). $r\simeq
Q_3/\sqrt{12 Q_4} \simeq 0.3$ ($n\sim -1.5$).  No significant time
dependence is found. The 
 corresponding cross correlation coefficients in multipole space are: 
\begin{equation}
Corr(l,\phi)=\frac{C_{SZ,G}(l,\phi)}{\left[C_{SZ}(l)C_G(l,\phi)\right]^{1/2}} 
\label{eqn:corr}
\end{equation} 

We now use the photometric redshift information from the galaxy survey
to vary the selection function $\phi$.  We pick the weighting which
maximizes (\ref{eqn:corr}).  This allows the measurement of one number
in the SZ to yield a full function of redshift $z$.  Since
(\ref{eqn:corr}) depends on one variable $l$, we can in principle
measure an optimal redshift weighting $\phi(z)$ at each $l$.  The
cross correlation variation has allowed us to measure a two
dimensional cross-correlation function from a one dimensional
observable in the SZ and a two dimensional observable in the galaxies.

\begin{figure}
\figurenum{7}
\plotone{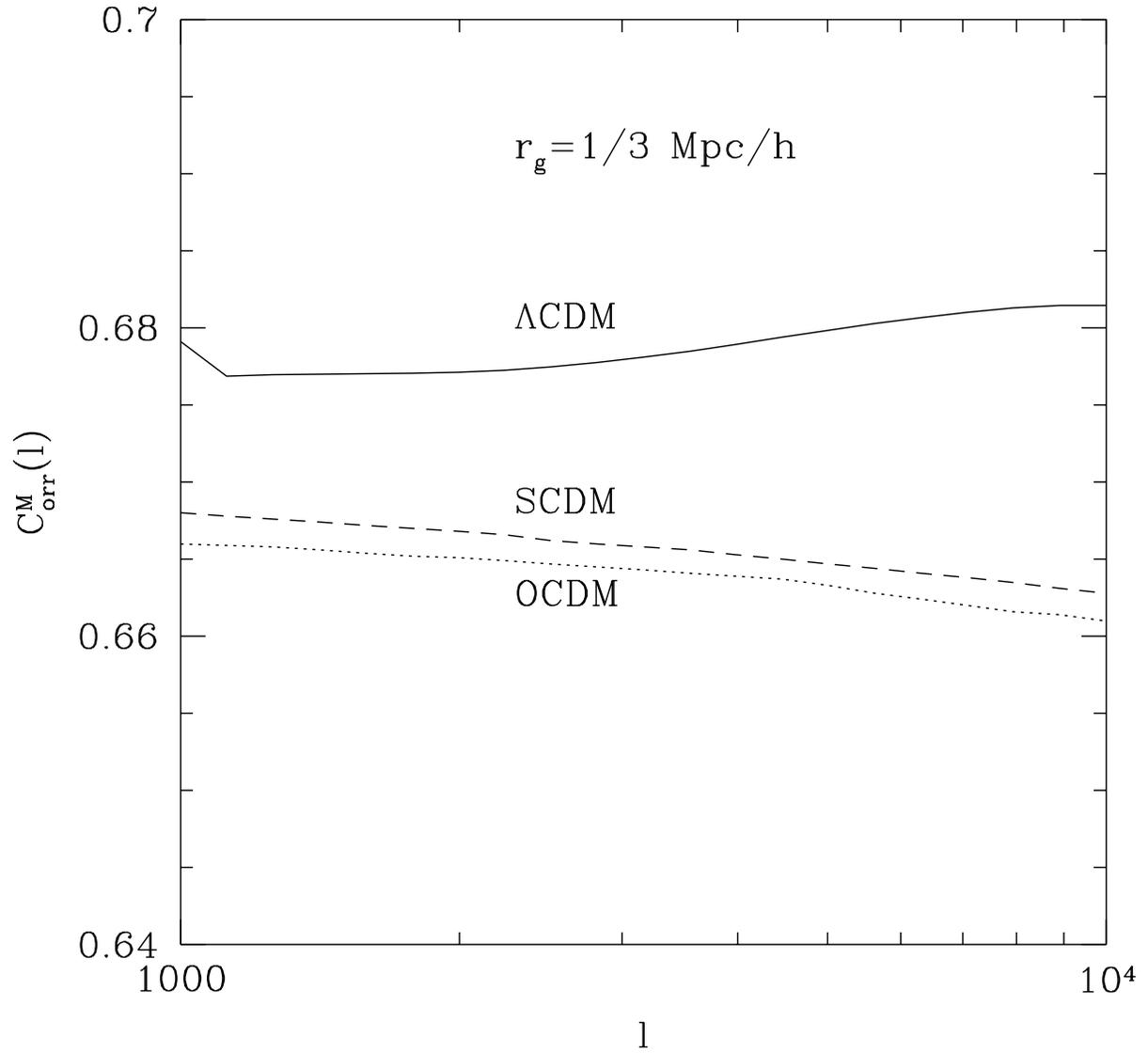}
\caption{SZ-Galaxy cross correlation coefficients in multipole 
space. Results have almost no dependence on cosmological
models and gas model parameters. \label{fig:Corr}}
\end{figure}

The selection function to maximize  $Corr(l)$
is obtained from the variation $\frac{\delta Corr(l,\phi)}{\delta
\phi}=0$. We 
denote this selection function as $\phi_M(l,z)$ and find that
\begin{equation}
\phi_M(l,z)=\alpha(l) \frac{\Delta^2_{SZ,G}(k,z)}{\Delta^2_G(k,z)}|_{k=l/x(z)}
\label{eqn:phim}
\end{equation}
Here, $\alpha(l)$ is a constant to be determined later by the
observational data.
The optimized $C^Morr$ is:
\begin{equation}
C^Morr(l)=\left[\frac{\int_0^{z_G} \Delta^2_{SZ,G} \phi_M C(x)
x(dx/dz)dz}{\int_0^{z_{cmb}} \Delta^2_{SZ} C(x) x(dx/dz) dz}\right]^{1/2}=
\left[\frac{\int_0^{z_G} \frac{\Delta^4_{SZ,G}}{\Delta_G^2}C(x)
x(dx/dz)dz}{\int_0^{z_{CMB}} \Delta_{SZ}^2 C(x) x(dx/dz)dz}\right]^{1/2}
\end{equation} 
Results are shown in figure \ref{fig:Corr}. Following the same 
estimation as in $r(k,z)$, $C^Morr(l) \simeq 3/4 \times
Q_3/Q_4^{1/2}\simeq 0.7$. The observed $C^Morr(l)$ may be smaller
because of the 
limited galaxy survey depth. 

The observationalrocedure to extract the redshift information is as
follows. 1. Start with a random guess for $\phi(z)$,
e.g. $\phi=1$ . 2. Given a photometric galaxy survey and SZ survey,
measure $C_{SZ}(l)$, $C_G(l,\phi)$ and 
$C_{SZ,G}(l,\phi)$  from angular correlation functions $\langle
\Theta(\hat{n}) \Theta(\hat{n}+\hat{\theta}) \rangle$,  $\langle
\frac{\Delta n}{n}(\hat{n}) \frac{\Delta n}{n}(\hat{n}+\hat{\theta})
\rangle$ and  $\langle 
\Theta(\hat{n}) \frac{\Delta n}{n}(\hat{n}+\hat{\theta})
\rangle$, respectively. 3. Vary $\phi$ to maximize $Corr(l)$ for a
specific $l$,
therefore obtain $\phi_M(l,z)$. 
4. Apply Eq.(\ref{eqn:phim}) to infer
$\Delta^2_{SZ,G}(l/x(z),z)$, up to a constant $1/\alpha(l)$,  from the
directly measured $\Delta^2_G$ from 
the galaxy survey.  Furthermore, Eq. (\ref{eqn:szg}) enables us to infer
$\alpha(l)$ from the observed $C_{l,obs}^{SZ,G}$.
\begin{equation}
\label{eqn:alpha}
\alpha(l)=\left[\frac{2 \times 16 \pi^2}{(2l+1)^3}\int_0^{z_G}
 \phi_M^2(l,z) \Delta_G^2[l/x(z),z] C(z) x(z)dx/dz dz
 \right]/C_{l,obs}^{SZ,G}
\end{equation}  5. Combine Eq.(\ref{eqn:rkz}) and Eq.(\ref{eqn:phim})
 to obtain  
the time resolved SZ power spectrum, up to a
factor $r(l/x(z),z)^{-2}$.
\begin{equation}
\Delta^2_{SZ}(k,z)=\phi_M^2(l,z)
\Delta^2_G(l/x(z),z)/[r^2(l/x,z)  \alpha^2(l)]
\end{equation}  In our model $r(l/x(z),z)$ is
almost a constant $\simeq 
1/0.7^{2}\simeq 2$ independent of cosmological models or redshifts (see Figure
\ref{fig:r}). This gives a theoretical estimate of
$\Delta^2_{SZ}(l/x,z)$.  6. $1/r^2$    
nomalizes the observed SZ anisotropy multipole $C_l^{obs}$ and the SZ
temperature variance. Apply Eq. (\ref{eqn:cl}) to integrate
$\Delta^2_{SZ}$ obtained above with $1/r^2=1$ and  compare
with $C_l^{obs}$, we will obtain the averaged $1/r^2$. 
\begin{equation}
\label{eqn:barr}
\langle \frac{1}{r^2(l/x(z),z)} \rangle=C_l^{obs}/\left[\frac{64
\pi^2}{(2l+1)^3} 
\int \phi_M^2(l,z) \Delta^2_G[l/x(z),z]/\alpha^2(l)C(x)x(z)dx/dz dz \right]
\end{equation} 7. Follow the
same steps to obtain $\Delta^2_{SZ,G}(l/x(z),z)$ and
$\Delta^2_{SZ}(l/x(z),z)$ at different $l$. 

Since for $C_{SZ}(l)$ at 
different angular scale $l$, $\langle 1/r^2(l/x,z)\rangle$ is determined
roughly by 
$r(l/x(z_p),z_p)$ ($z_p\sim 1$ is the redshift with peak
contribution to $C_l$. See Fig. \ref{fig:zcon}),  we can even get
some idea about the scale dependence of $r(k,z)$. Thus, the total 
projected SZ autocorrelation give a consistency check on  
the reconstructed time resolved power spectrum from the galaxy-SZ
cross correlation. Furthermore,  the
galaxy bias and its time dependence have completely dropped out of the
calculation, and are thus not expected to affect the results at all.
Then, in principle, SZ-galaxy correlation plus 
SZ CMB anisotropy provide a consistent and powerful method  to extract
all time evolution information of the IGM pressure power spectrum.  
  
Noise and cosmic variance put constraints on the feasibility of our
procedure. (1) Limitation of CMB resolution degrades our method. The
measured range of $k$ is 
$[l_1/x(z),l_2/x(z)]\sim[1_1/3000z,l_2/3000z]
h$/Mpc. Here, $[l_1,l_2]$ is the range of the CMB experiment. In order
to detect the peak of $\Delta_p^2$ ( around $k= 
3h$/Mpc as shown in Fig. \ref{fig:pressure}),  $z\leq l_2/9000 $. For CBI ($630\leq
l\leq 3500$), we are only able to detect $z\leq 0.4$.  AMIBA will
measure $l\leq 28500$ and \citet{10m} will
measure  $l \leq 40000$. They will allow us 
to measure the gas power spectrum  up to $z\sim 3$ and $z\sim 4$, respectively.
(2) Observational errors impose further
constraints. Suppose that the  galaxy 
survey covers a fraction $f_G$ of the sky and the $i$-th survey region
(For example, if the redshift accuracy is $\Delta z$, then we can
divide the galaxies into redshift bins with $0 \leq z \leq \Delta z$,
$\Delta z \leq z \leq 2 \Delta z $, etc. ) have
$N_i$ observed galaxies. The CMB observation covers 
a fraction $f_{cmb}$ of sky and the $C_l$ is averaged over the
band $[l-\Delta l/2,l+\Delta l/2]$. Then the galaxy number count
causes  the Poisson error:
\begin{equation}
\frac{\Delta C_G}{C_G}\sim \frac{\Delta
C_{SZ,G}}{C_{SZ,G}}\sim \left[f_G \times min(N_i)\right]^{-1/2}
\end{equation}
 (3) The cosmic variance of the $C_l$ also cause errors. Recalling that
$C_l=\sum a_{lm}a_{lm}^*/(2l+1)$ and 
$a_{lm} \propto \delta T \propto b_p \delta$, we get:
\begin{equation}
\frac{\Delta C_l}{C_l}=\sqrt{\frac{\langle C_l^2\rangle-\langle C_l
\rangle^2}{\langle C_l 
\rangle^2}} \sim \sqrt{\frac{S_4 \sigma_R^2}{(2 l+1) \Delta l
f_{cmb}}}  \sim \sqrt{\frac{1}{10 (l/2000) \Delta l f_{cmb}}}
\end{equation}  
Here, $\sigma^2_R$ is the density dispersion over smoothing scale
$R\sim h/$Mpc. We already use the typical value of $S_4
\sim 40$ and  $\sigma^2_R(z=1) \sim 10$. The corresponding error caused in
$\phi_M$ is:   
\begin{equation}
\frac{\Delta \phi}{\phi_M}\sim
\sqrt{\frac{\Delta Corr}{\phi^2_M \frac{\delta^2 Corr}{\delta
\phi^2}|_{\phi_M}}} \sim \left[(10 (l/2000) \Delta l
f_{cmb})^{-1}+f_{G}^{-1} \times min(N_i)^{-1}\right]^{1/4} 
\end{equation}
Recalling that $\Delta^2_{SZ} \propto  \phi_M^2$, requiring a $40\%$ accuracy
on $\Delta^2_{SZ}$ would impose that (a) $f_G \times
min(N_i)\geq 10^3$. Each survey regions must be large 
enough in order to contain sufficient number of galaxies and must be
small enough to ignore the evolution. We may choose redshift bands
of each survey region $\Delta z_i\sim 0.1$. Then the number of
galaxies observed $N_O$ has to satisfy
$N_O \geq 10^3 z_G/\Delta z_i/f_G$. For SDSS (\citet{SDSS}),
which covers one quarter of the sky and probes more than  one million
galaxies with photometric redshift up to $z\sim 1$, the requirement
$N_O \geq 10^5$ is easily satisfied. The measurement of the
intergalactic gas at redshift z requires the galaxy survey at least up
to that redshift (equation \ref{eqn:phim}), so we need deeper galaxy
survey in order to probe the gas beyond $z\sim1$. (b)
$f_{cmb} \Delta l \geq  10^2 (2000/l)$. For CMB experiments with 
relatively lower resolution, larger sky coverage is required.
For example, though Planck only measures $l\leq 2000$, it covers the
whole sky and therefore satisfies this condition. For those with much
higher resolution such as AMIBA and Submillimeter Telescope, the
required sky coverage can be relaxed to the order of $1 \%$.

This variation method
only depends on the assumption that the cross correlation coefficient
$r(k,z)$ is approximately a constant, which is the direct result of
the hierarchical model and has only weak
dependence on the gas  model and cosmologies.  Furthermore, the
hierarchical model is strongly supported by the consistency of the CMB
SZ power spectrum dependence on $\sigma_8$ and the behavior of the gas bias
between our model and simulations. Since the averaged $r(k,z)$ is
measurable, our method does not rely much on the theoretical
value of $r(k,z)$  and thus observationally consistent.

\section{Non-gravitational heating effect}
\label{sec:NGheating}
Pen (1999) has shown that the IGM has most likely been preheated by
non-gravitational energy sources with  energy injection $E_{NG}\sim 1$
keV per nucleon. This section is devoted to consider this
effect. Because the relation between the gravitational
heating and the  non-gravitational heating is very uncertain, we only consider two extreme cases. The 
first one is that the non-gravitational heating is perfectly correlated
with the gravitational heating, then we can change 
Eq. (\ref{eqn:temp}) to: 
\begin{equation}
\label{eqn:corrNGT}
kT_g=\frac{1}{6} (1+\beta) \frac{4 m_H}{3+5 X} (\Psi-\bar{\Psi})   
\end{equation}
We will append all
former results from gravitational heating with a superscript 'A'
(Adiabatic). Here, $\beta\equiv T_{NG}/\bar{T_g}^A$ represents the ratio of the
non-gravitational heating and the gravitational heating.
$kT_{NG}=8/3/(3+5 X) E_{NG}$. $E_{NG}\sim 1$ keV
corresponds to $\beta \sim 
1$. All former results are not affected by this change except
$\bar{T_g}$, $\bar{y} \propto \bar{T_g}$,
$C_l \propto \bar{T_g}^2$ and $\Theta_3 \propto 1/\bar{T_g}$ due to
the dependence $\bar{T_g} \propto (1+\beta)$. Figure \ref{fig:temp} and
\ref{fig:cl} need to be changed correspondingly. Figure \ref{fig:zcon}
and \ref{fig:Corr} will change only when $\beta$ is
time-dependent. All other figures remain the same.

The second case is that non-gravitational heating is uncorrelated
with local density, then, Eq. (\ref{eqn:temp}) changes to:
\begin{equation}
\label{eqn:uncorrNGT}
kT_g=\frac{1}{6} \frac{4 m_H}{3+5 X} (\Psi-\bar{\Psi})+kT_{NG}(z)
\end{equation}
$\bar{T_g}=\bar{T_g}^A+T_{NG}=[1+\beta(z)]
\bar{T_g}^A$.  Then, the
corresponding new results following the same definitions are:
\begin{eqnarray}
\label{eqn:newpressure}
P_p(k,z)&=&\left[P^{A}_p(k,z)+(2 \pi)^3 \delta_D({\bf k}) (2
\beta+\beta^2)+2 \beta P^{A}_{SZ,\delta}(k,z)+\beta^2
P(k,z)\right]/(1+\beta)^2 \nonumber \\
 &=&\frac{(b_p^A)^2+2 \beta r^{A}b_p^{A}+\beta^2}{(1+\beta)^2} \times P(k,z)
\ ({\rm when} \ k\neq 0)
\end{eqnarray}
\begin{equation}
\label{eqn:newSZG}
P_{p,G}(k,z)=\frac{b}{1+\beta} \times
\left[P^{A}_{p,\delta}(k,z)+\beta P(k,z)
\right]=b\times\frac{r^{A}b^{A}_p+\beta}{1+\beta} P(k,z) 
\end{equation}
\begin{equation}
\label{eqn:newbias}
b_p(k,z)=\frac{\sqrt{(b_p^A)^2+2 r^{A} b^{A}_p \beta+\beta^2}}{1+\beta}
\end{equation}
\begin{equation}
\label{eqn:newr}
r(k,z)=\frac{r^{A} b_p^{A}+\beta}{\sqrt{(b^A_p)^2+2
r^{A} b^{A}_p \beta+\beta^2}}
\end{equation}
Here, $P_{p,\delta}^A=P^A_{p,G}/b$ is the Fourier transform of
$\langle y_p^A({\bf x}) \delta({\bf x+r})\rangle$. 
The expression of the optimal selection function $\phi_M$
(Eq. \ref{eqn:phim}) is not affected at all. Equation
(\ref{eqn:newr}) tells us that  $r^A \leq r \leq 1$.
As expected, when $\beta \ll b_p^A$, we reduce to our former
results and when $\beta \gg b_p^A$, we obtain the bias model. There
are three regions: (a) $k
\gg 1/r_g \sim 3h $/Mpc. We always have $\beta \gg b_p^A$, $b_p
\rightarrow \beta/(1+\beta)$ and $r\rightarrow 1$. Our redshift
extraction method works well with $r\simeq 1$ in this region. Because
this region is dominated by non-gravitational heating,  high
resolution CMB experiments are able to measure $\Delta^2_{SZ}$ in this
region and  enable us to directly obtain the information of the
non-gravitational heating. Since $b_p$ 
does not approach zero as before, $\Delta^2_{SZ}\propto \Delta^2$ increases
with increasing $k$. This behavior will move the peak of SZ power
spectrum to larger $l$ while increasing the amplitude of SZ power
spectrum. (b) $k \lesssim 6h$/Mpc.  Then,  $b_p^A \sim 8 \gg \beta(0)\sim1$, so $b_p(z)$ will
decrease roughly by a factor $(1+\beta(z))$. The
resulting $r$ is bigger than our former results, but it still remains
roughly  a constant with respective to space and time, so our
redshift extraction method works well. (c) The intermediate
region ( the range 
$b_p^A \sim \beta$, roughly $6 {\rm h/Mpc} \lesssim k \lesssim
10{\rm h/Mpc}$.), $r$ varies from
$\sim 0.7$ to 1. Our redshift deprojection method still works, but
with a larger error. The discussion of the observational requirement is not
affected  at all. The new $Corr$ should
differ only a little bit from the former result due to its
dimentionless definition. Roughly, $C_l \propto (1+\beta)^2$,
bispectrum $\propto (1+\beta)^3$ and skewness $\Theta_3\propto (1+\beta)^{-1}$.

\begin{figure}
\figurenum{8}
\plotone{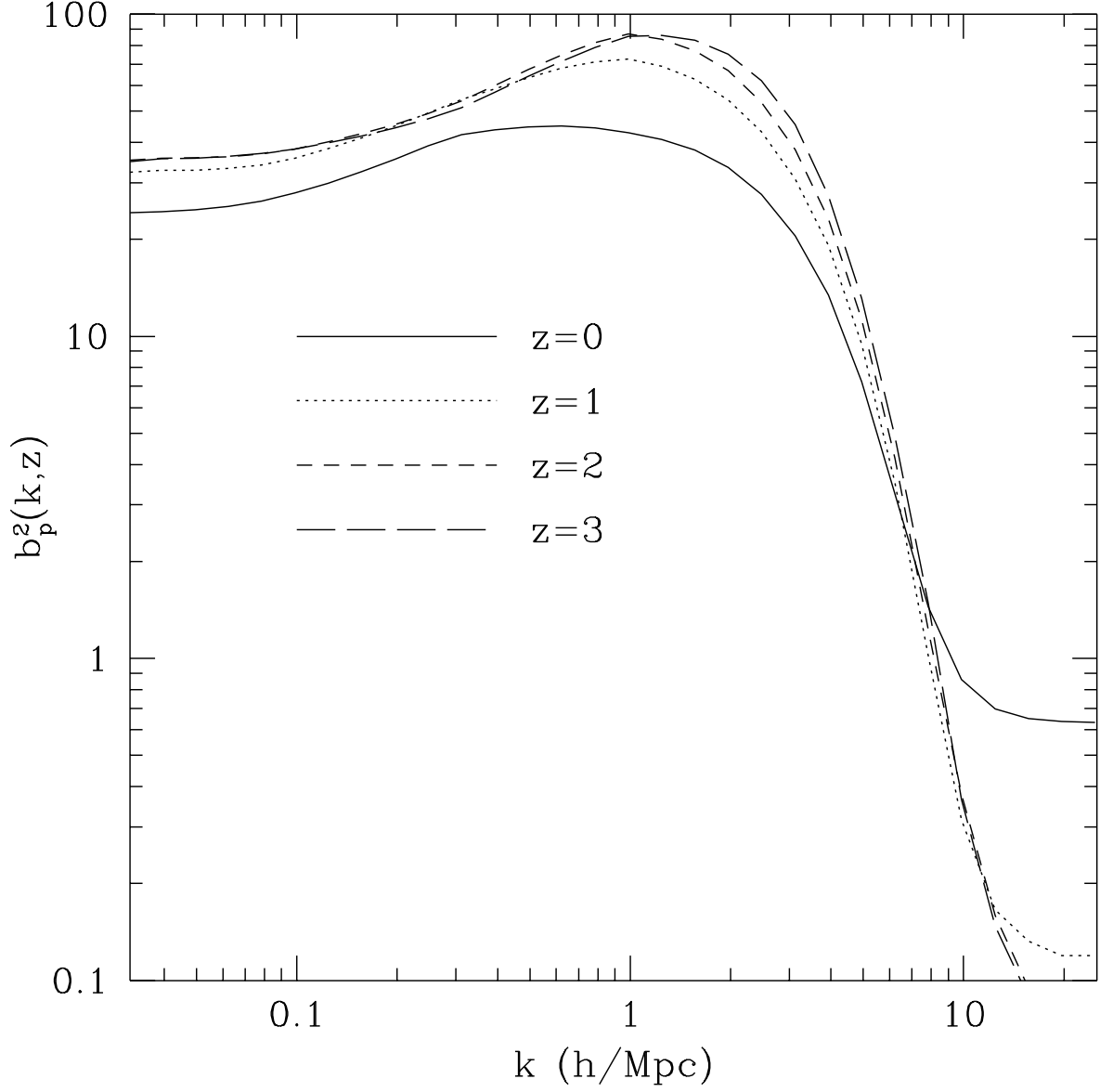}
\caption{$b_p^2(k,z)$ and $\Delta^2_p(k,z)$ when non-gravitation
heating is considered. $kT_{NG}=0.39 {\rm keV} (1+z)^{-3}$ is  assumed. The
time evolution of $b_p^2$ is partly due to the 
evolution of $\beta$. 
\label{fig:newbias}} 
\end{figure}

\begin{figure}
\figurenum{9}
\plotone{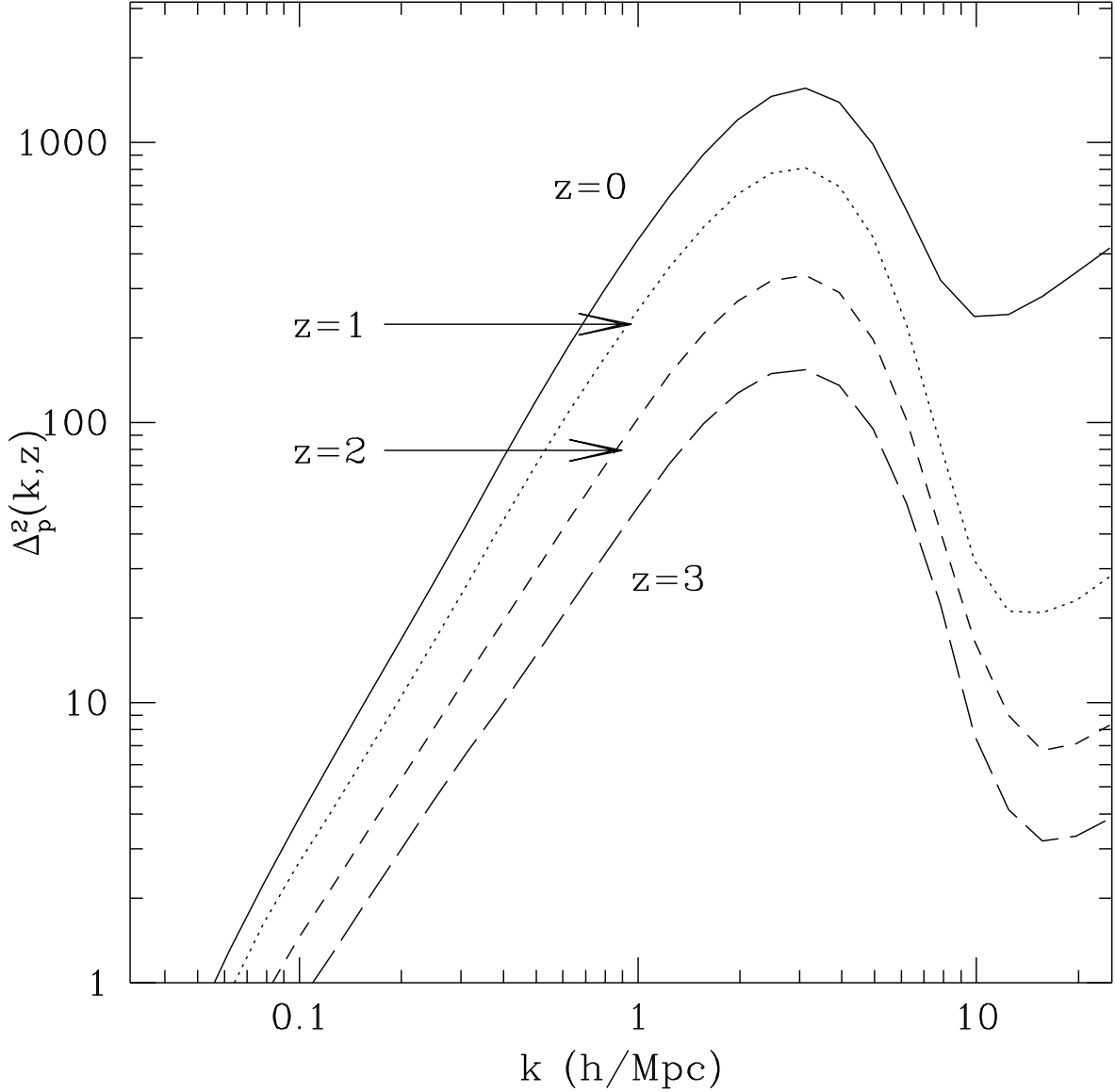}
\caption{$\Delta^2_p(k,z)$ when non-gravitation
heating is considered. $kT_{NG}=0.39 {\rm keV} (1+z)^{-3}$ is
assumed. The distinctive behavior of $\Delta^2_p$ at
small scales ($k \gtrsim 10 \rm{h/Mpc}$) is due to the
non-gravitational heating and the measurement of this region will help
us to extract the non-gravitational heating information.
\label{fig:newpp}} 
\end{figure}

\begin{figure}
\figurenum{10}
\plotone{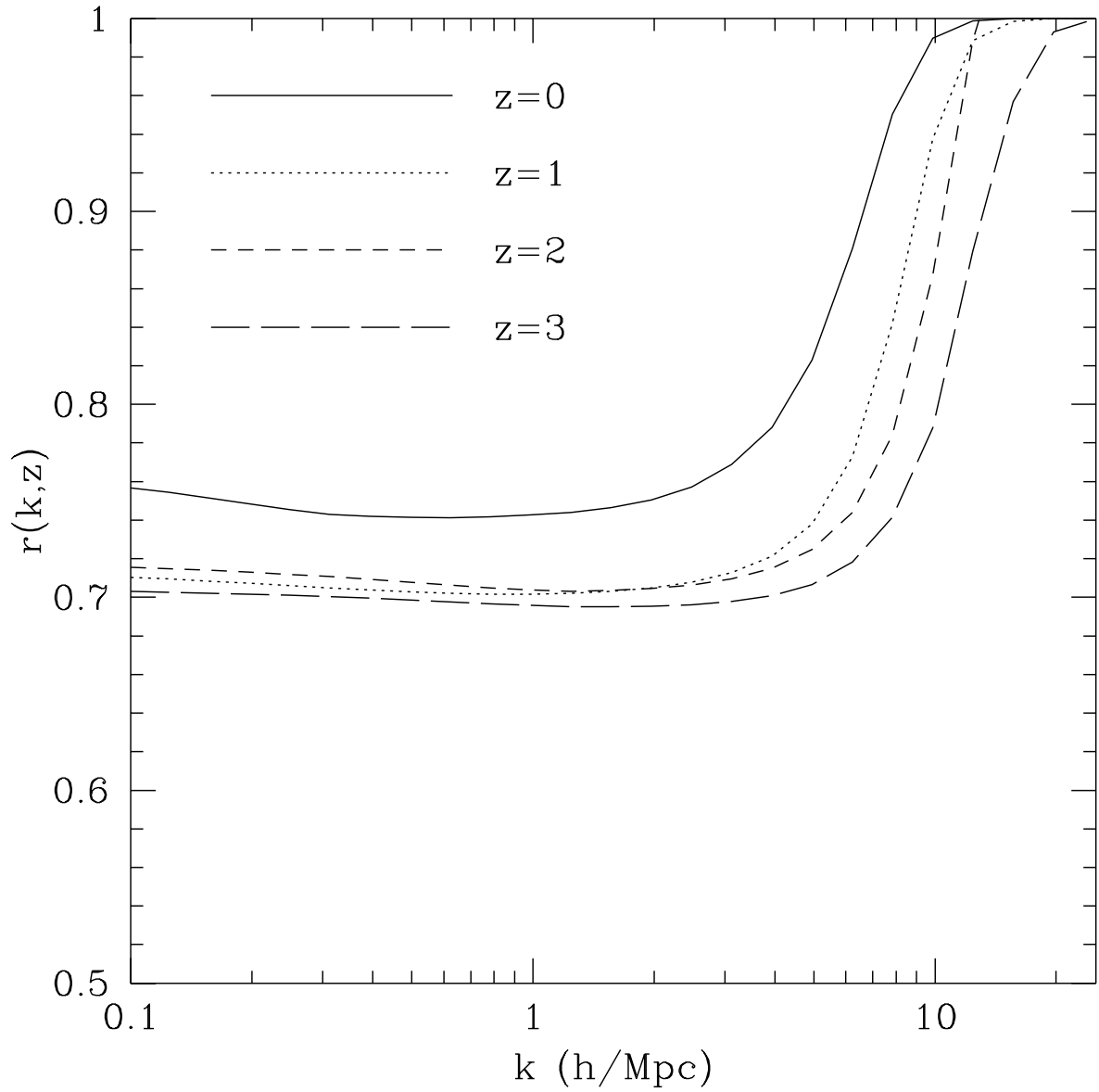}
\caption{$r(k,z)$ when non-gravitational heating is considered with
same assumptions as in figure \ref{fig:newbias}.}.
\label{fig:newr}
\end{figure}

Our knowledge on the non-gravitational heating is very limited, and
presumably depends on the poorly understood 
physics of star formation and supernovae dynamics.  It is not even
known if the heating was pre or post structure formation.  We have
surveyed two simple but different models of non-gravitational heating
to demonstrate the range of effects it may have on the
correlations. The above equations  show the basic procedure to include
the non-gravitational heating and the possible effects. So we do
not plan to constrain to a highly hypothetical model and do the
calculation. For the
purpose of estimation, we may either assume a 
step function for $T_{NG}$  ($T_{NG}(z)=T_{NG}(0)$ for $z<z_{NG}$. Otherwise,
$T_{NG}=0$.) or  $T_{NG}(z)=T_{NG}(0) (1+z)^{-\gamma}$.
The upper limit of '$y$' parameter $y\leq 1.5\times 10^{-5}$ from
COBE/FIRAS \citep{Fixsen96} put constraints on the value of $z_{NG}$
and $\gamma$.  For example, when $T_{NG}=1$ keV, $\gamma >1.5$ or
$z_{NG} < 3 $. Further simplification can be made by
taking the fitting formula of  $(b^A_p)^2$  and
$r^A\simeq\bar{r}^A\simeq 0.7$.  We show some examples (Figure
\ref{fig:newbias},\ref{fig:newpp} and \ref{fig:newr}).

When combining CMB experiments with galaxy surveys, we have  four observables:
$C_l$, $\Delta^2_G(k,z)$, $\Delta^2_{SZ,G}(k,z)=\phi_M(k,z)
\Delta^2_G(k,z)$ and $\bar{r}(k)$. If  $r(k,z)$ is not a strong
function of time and scale, we can solve for $\Delta^2_{SZ}(k,z)$. If
the behavior of 
$b_p(k,z)$ as discussed  above is correct, we obtain the fifth observable
$\beta(z)$ from the small scale behavior of $\Delta^2_{SZ}$.
    
\section{Discussion}
\label{sec:discussion}
Let us now address some potential systematic shortcomings in our
simplified model.  (1) Our model relies on the adopted gas-dark matter
density relation (Eq. \ref{eqn:gasw}), the (non-local)
temperature-density relation (Eq. \ref{eqn:temp}) and the
non-gravitational heating. In case that only the gravitational heating is
included,  apart from intrinsic
problems addressed in section \ref{sec:ps}, our model contains two free
parameters: the gas-dark matter smoothing length, and the
temperature-potential smoothing length.  These parameters only weakly affect
the pressure 
power spectrum and pressure bias, which are defined in terms of
dimensionless functions.  The normalization is expressed in terms of
the mass weighted temperature, $\bar{T_g}$. We  see in
Table \ref{tble:tbar} that the gas smoothing only has a small effect.
A preheating of 1 keV as proposed by \citet{Pen99} has a factor of
order unity
impact on the temperature, and about factor of 4 effect 
on the CMB  anisotropy spectrum and 8 on the bispectrum, while the
skewness will decrease by about 50\%. We provide a
convenient method to include such effects.
Hydro simulations are able to measure the gas-DM
relation and the $T-\Psi$ relation and will further improve our work.
Fortunately, the main goal of our model---the extraction 
of the 3-D gas information is least affected, since it depends only on
the cross-correlation coefficient (\ref{eqn:rkz})
and Figure \ref{fig:r}, which does not vary a lot even including
non-gravitational heating.
(2) The galaxy bias model we adopt may be too simple. In our redshift
extraction method of the gas pressure power spectrum, the only bias
dependency comes from $r(k,z)$. We speculate that, even for a realistic galaxy
bias model, $r(k,z)$ should be close to a constant, which still enable
us to extract $\Delta^2_{SZ}$.
(3) $Q_N^{sat}$. The value of $Q_N^{sat}$ in HEPT is a function of the
power index of  a power law power spectrum. We have extended it to the
CDM power spectrum and  choose the power index at $\sum k_i/N$.
Because SZ effect is mainly contributed by the non-linear regime with
$n \sim -1.5$, HEPT is applicable and the resulting $Q_N$ does not vary
a lot. More importantly, $Q_3/Q_4^{1/2}$ only has a weak dependence on
$n$. This behavior ensures that $r(k,z)$ has least dependence on our
assumptions about $Q_N^{sat}$.
\section{Conclusions}
\label{sec:conclusion}

We have presented a new tool to compute power spectra and other
statistics of SZ fluctuations based on hierarchical clustering and
scaling.  This approach describes the two point correlation of
non-linear observables, such as the SZ $y$ parameter in terms of two
point correlation functions, which directly maps to the observed
angular power spectrum.  We have demonstrated that this approach is
feasible, and produces results consistent with simulations and
Press-Schechter approaches.  We then addressed the problem of
measuring the redshift evolution of the SZ effect through cross
correlation with galaxy surveys containing coarse photometric redshift
information.  We have shown that a variational method allows the
redshift deconvolution which does not depend on galaxy biasing.  The
only model dependent quantity is the cross correlation coefficient $r$
between galaxies and gas pressure, whose value has least dependence on
the gas density model and the gas temperature model.

Our quantitative estimates suggest that, when combining CMB
experiments either with high resolution such as AMIBA and South Pole
Submillimeter Telescope or with high sky coverage such as Planck and
deep broad photometric galaxy surveys such as Sloan, our method is
able to extract the redshift evolution information of intergalactic
gas, such as the full time resolved gas pressure power spectrum, even
without requiring the knowledge of galaxy bias. It is even capable 
of disentangling the contribution from the gravitational heating and
those from the non-gravitational heating. This method serves as a
powerful probe to this primary component of cosmic baryons to high
redshift.  The model also provides an 
alternative to simulations and the Press-Schechter formalism to
calculate the CMB SZ power spectrum and bispectrum.  We successfully
reproduce, when no non-gravitational heating presents, the mass
weighted gas temperature $\bar{T_g} (\sim0.35$ keV), 
the pressure power spectrum, the pressure bias ($\sim 8$, but scale
dependent), the mean SZ temperature distortion ($\sim 5\times 10^{-6}$
K), the SZ power spectrum and the skewness parameter ($\sim
10^6$). With our transform formulas, non-gravitational heating is
easily included and we have estimated its effects on the correlations.



\end{document}